\documentclass[aps,prl,twocolumn,superscriptaddress]{revtex4}
\usepackage{graphicx}
\usepackage{latexsym}
\usepackage{amssymb}
\usepackage{amsmath}
\usepackage{amsfonts}
\usepackage{bm}
\usepackage{multirow}
\usepackage{color}
\usepackage{hyperref}

\newcommand{\ii}{\mathrm{i}}

\newcommand{\bra}[1]{\langle #1|}
\newcommand{\ket}[1]{|#1 \rangle}
\newcommand{\braket}[1]{\langle #1 \rangle}
\newcommand{\dsZ}{\mathbb{Z}}
\newcommand{\Tr}{\mathrm{Tr}}

\newcommand{\mat}[1]{\left(\begin{smallmatrix}#1\end{smallmatrix}\right)}

\newcommand{\refcite}[1]{Ref.\,\cite{#1}}
\newcommand{\eqnref}[1]{Eq.\,(\ref{#1})}
\newcommand{\figref}[1]{Fig.\,\ref{#1}}

\newcommand{\secref}[1]{Sec.\,\ref{#1}}
\newcommand{\appref}[1]{Appendix~\ref{#1}}

\newcommand{\beq}{\begin{equation}}
\newcommand{\eeq}{\end{equation}}
\newcommand{\beqn}{\begin{align}}
\newcommand{\eeqn}{\end{align}}

\hypersetup{
  colorlinks = true,
  urlcolor = blue,
  pdfauthor = {Kevin Slagle, Yi-Zhuang You, Cenke Xu},
  pdftitle = {Disordered XYZ Spin Chain Simulations using SBRG}
}

\newcommand{\J}{\tilde J}
\newcommand{\SE}{S_\text{E}}

\newcommand{\footError}{(error bar details: \cite{foot:errors})}

\begin{document}

\title{Disordered XYZ Spin Chain Simulations using the \\
Spectrum Bifurcation Renormalization Group}

\author{Kevin Slagle}

\author{Yi-Zhuang You}

\author{Cenke Xu}

\affiliation{Department of physics, University of California,
Santa Barbara, CA 93106, USA}

\begin{abstract}

We study the disordered XYZ spin chain using the recently
developed Spectrum Bifurcation Renormalization Group (SBRG)
\cite{You:2015sb} numerical method. With strong disorder, the
phase diagram consists of three many body localized (MBL) spin
glass phases. We argue that, with sufficiently strong disorder,
these spin glass phases are separated by marginally many-body
localized (MBL) critical lines. We examine the critical lines of
this model by measuring the entanglement entropy and
Edwards-Anderson spin glass order parameter, and find that the
critical lines are characterized by an effective central charge
$c'=\ln2$. Our data also suggests continuously varying critical
exponents along the critical lines. We also demonstrate how
long-range mutual information introduced in \refcite{Qi:2015mi}
can distinguish these phases.

\end{abstract}


\maketitle

\section{Introduction}

Quantum phase transitions \cite{Sachdev:2011lq} were previously
discussed as transitions between ground states of quantum
many-body systems at zero temperature. %The conventional wisdom is
%that quantum effects are only prominent for ground states and a
%few low-lying excited states in the many-body spectrum.
The conventional wisdom is that highly excited states of a
many-body
system at finite energy density are typically %decoherent and
self-thermalized following the eigenstate thermalization
hypothesis (ETH)
\cite{Deutsch:1991ik,Srednicki:1994ns,Rigol:2008qr}. However,
later it was realized that many-body localized (MBL) systems
\cite{GMP:2005,BAA:2006,Huse:2007is,Znidaric:2008he,Imbrie:2014,Huse:2015rv}
-- typically systems with quenched disorder -- can evade
thermalization and exhibit robust quantum coherence and
non-ergodic dynamics even at finite energy density. %\footnote{See
%\appref{sec: MBL} for a brief review of MBL and marginal MBL
%systems.}
The phenomenon of many-body localization (MBL) enables quantum
phase transitions to occur at a finite energy density between
different MBL quantum phases
\cite{Refael:2013du,Huse:2013ci,Altman:2014hg,Chandran:2014fv,Potter:2015th,Xu:2015gg}.
The corresponding quantum critical points are marginally localized
\cite{Potter:2014mh} and are thus known as marginal MBL states (or
quantum critical glasses) \cite{Vasseur:2015qf}. Similar to MBL
states, marginal MBL states are also non-ergodic,
%with no thermal conductivity,
and can be specified by an extensive number of
quasi-local integrals of motion (LIOM)
\cite{Abanin:2015io,Kim:2014zj,Scardicchio:2015kl}.
%However, the quasi-LIOM of the marginal MBL states are not
%exponentially localized as those of the MBL states. Instead, they
%are power-law localized and exhibit scale-invariant behaviors
%resembling the zero-temperature quantum critical systems,
Unlike the typical MBL states, the marginal MBL states exhibit
critical behaviors, including the logarithmic scaling of
entanglement entropy (in 1D) and the power-law decay of
disorder-averaged correlation functions and mutual information.

In this work, we will study the marginal MBL states in a 1D XYZ
spin model using the spectrum bifurcation renormalization group
(SBRG) numerical method introduced in \refcite{You:2015sb}. SBRG
is another version of the excited-state real space renormalization
group (RSRG-X)
\cite{Altman:2013rg,Altman:2014hg,Refael:2013du,Potter:2015th},
which is specifically designed for the class of MBL models that
has a bifurcating spectrum branching structure at each
renormalization group (RG) step. The idea of SBRG is similar to
RSRG-X, which targets the full many-body spectrum rather than just
the ground state. Given a many-body Hamiltonian $H$ with strong
disorder, at each RG step, the leading energy scale term  (the
strongest local term) $H_0$ in the Hamiltonian $H$ is first
selected, and the whole Hamiltonian $H$ is rotated to the block
diagonal basis of $H_0$. The block off-diagonal terms, which
resonate between different $H_0$ eigen sectors, are treated as
perturbations and are reduced to effective terms within the
diagonal block via second order perturbation. The RG procedure
gradually block diagonalizes the many-body Hamiltonian until it is
fully diagonalized. \footnote{See \appref{sec: SBRG} for details
of the SBRG method.} The resulting effective Hamiltonian
$H_\text{eff}$ can be viewed as the RG fixed point Hamiltonian for
the MBL system
\cite{Abanin:2013lc,Abanin:2013ta,Abanin:2015io,Huse:2014ec,Kim:2014zj},
which encodes the full many-body spectrum. The RG
transformations can be collected and combined into a unitary
quantum circuit $U_\text{RG}$ (\figref{fig:holography critical},
\ref{fig:holography SG}), which encodes the matrix product state
(MPS) approximations for all eigenstates. Various physical
properties of the MBL (and marginal MBL) system can be calculated
based on the data of $H_\text{eff}$ and $U_\text{RG}$ generated by
SBRG. Unlike RSRG-X, SBRG does not explicitly choose a specific
eigen sector at each RG step. Instead, the spectrum branching is
encoded implicitly in the flow of the Hamiltonian, such that the
entire spectrum is targeted during each RG flow.

There has been a long history of using the real space
renormalization group (RSRG) method to study disordered spin
chains.\cite{Dasgupta:1980so,Lee:1982jv,Fisher:1992is,Fisher:1994he,Fisher:1995cr}
RSRG was originally proposed as a ground state targeting approach,
and has been applied to the random Heisenberg,\cite{Dasgupta:1980so} transverse field Ising,\cite{Fisher:1992is,Fisher:1995cr} XY and XXZ \cite{Fisher:1994he}
spin chains. It was found that ground states of (clean) quantum critical
spin chains (e.g. Ising, XX or Heisenberg) could be unstable to
random exchange couplings and flow to the infinite randomness
(strong disorder) fixed-point. Whether the critical phenomena of
the infinite randomness fixed-point persists at finite energy
density is further discussed in the context of MBL using RSRG-X
and other methods
\cite{Altman:2014hg,Potter:2015th,Refael:2013du,Vasseur:2015ys}.
The current understanding is that the strong disorder criticality
could persist to finite energy density as marginal MBL states in
quantum Ising chains. \cite{Altman:2014hg} But for (planar) XXZ
and Heisenberg chains, the marginal MBL state is unstable towards
thermalization or spontaneous symmetry breaking
\cite{Potter:2015th,Vasseur:2015ys}, due to the extensive number
of local degeneracies dictated by the symmetry group. Take the
random XXZ chain for example, the symmetry group $U(1)\rtimes
\dsZ_2$ is a product of the spin-Z conservation and the
spin-flipping $\dsZ_2$ symmetry. Following the argument given in \refcite{Potter:2016}, the symmetric marginal MBL state,
if possible, should be characterized by a set of quasi-LIOM, which
form irreducible representations of the $U(1)\rtimes \dsZ_2$
symmetry group. However, the $U(1)\rtimes \dsZ_2$ symmetry
enforces a local degeneracy between states of opposite spin-Z for
every quasi-LIOM. As a result, the finite energy density
eigenstates in the many-body spectrum are all exponentially
degenerate. Because the extensive degeneracy is unstable to
quantum fluctuations, the eigenstates must either localize to
symmetry breaking spin glass states or thermalize, both of which
destroy the quantum criticality.

So to explore marginal MBL phases in 1D spin systems, we must
sufficiently break the symmetry to remove all local degeneracies.
This motivates us to look at the XYZ spin chain, with
independently random XX, YY and ZZ couplings on each bond. The
symmetry is broken down to $\dsZ_2^2$, such that local
degeneracies are completely removed (because $\dsZ_2^2$ has no
irreducible representations beyond the one-dimensional representations).
Therefore, marginal MBL states of the XYZ spin chain can be stable
at finite energy density against thermalization and spontaneous
symmetry breaking as long as the disorder is strong enough. Due
to the discrete $\dsZ_2^2$ symmetry, at each RG step there is only
one unique leading energy scale term that bifurcates the spectrum.
This is precisely the type of model that SBRG was designed to
target.

Apart from the above symmetry considerations, strong disorder is
another key ingredient to keep the marginal MBL states from
thermalizing. We introduce the standard deviation of the
logarithmic scale of the coupling strengths
\begin{equation}
  \Gamma=\mathrm{std}(\ln |J|)
  \label{eq:stdJ}
\end{equation} (where $J$
appears as coefficients in the Hamiltonian, e.g.
\eqnref{eq:model}) to compare the strength of three often used
disorder distributions: uniform, Gaussian, and the power-law
distribution which will be used in this work. Physically, $\Gamma$
describes how much different couplings are separated in their
energy scales. Well separbated energy scales in the large $\Gamma$
limit suppress the resonance between energy levels, and hence
hinders thermalization. Our finite-size exact diagonalization (ED)
study (\appref{sec:ED}) indicates that $\Gamma \simeq 1$ is not
sufficient to stabilize the marginal MBL phases in the XYZ model
against thermalization. Therefore, instead of drawing the coupling
strengths from uniform distributions ($\Gamma=1$) or Gaussian
distributions ($\Gamma \approx 1.1$), we need to take power-law
distributions (\appref{sec:disorder distribution}) whose $\Gamma$
can be tuned all the way to infinity.
%(Note, large $\Gamma$ in \eqnref{eq:stdJ} may not always guarantee strong disorder; see \appref{sec:estimating disorder}.)
We will typically take $\Gamma=4$ as the initial distribution in
our calculation. SBRG is well-suited to study such strong disorder
spin systems, as the SBRG algorithm is asymptotically accurate in
the large $\Gamma$ limit.

In the following, we will first introduce the model and present
the phase diagram. Then we focus on a high symmetry line in the
phase diagram, and investigate the MBL spin glass phase and
marginal MBL critical phase in detail. In particular, we
calculate the entanglement entropy, Edwards-Anderson
correlator and long-range mutual information. We found that
the marginal MBL critical line is characterized by an effective
central charge $c'=\ln 2$. Our data also suggest continuously varying exponents along the critical line.

\section{XYZ Spin Chain Model}

We study the XYZ spin chain with large disorder and periodic
boundary conditions. The Hamiltonian is given by \beq
\label{eq:model} H = \sum_{i=1}^L \sum_{\mu=x,y,z} J_{i,\mu} \;
\sigma_i^\mu \sigma_{i+1}^\mu. \eeq $\sigma_i^\mu$ with $\mu =
1,2,3$ are Pauli matrix operators on lattice site $i$ of a 1d
chain of length $L$. The couplings $J_{i,\mu} \in [0,J_\mu]$ are
randomly drawn from a power-law distribution
\begin{equation}
\mathrm{PDF}(J_{i,\mu}) = \frac{1}{\Gamma J_{i,\mu}}
\left(\frac{J_{i,\mu}}{J_{\mu}}\right)^{1/\Gamma},
\end{equation}
where $0 < \Gamma < \infty$ (see \eqnref{eq:stdJ}) controls the disorder strength (for
details see \appref{sec:disorder distribution}). Equivalently,
$J_{i,\mu}^{1/\Gamma} \in [0,J_\mu^{1/\Gamma}]$ is uniformly
distributed. For later convenience, we define
\begin{equation}
\J_\mu \equiv J_\mu^{1/\Gamma},
\end{equation}
and take $\J_\mu = \J_x, \J_y, \J_z$ as our primary tuning
parameters (see \appref{sec:tuning parameters} for an
explanation). We will be interested in the entire energy spectrum
of this model, as opposed to just the low energy states.

%This model has $\dsZ_2^2$ symmetry with the following three
%nontrivial elements: $\Pi^\mu = \prod_i \sigma^\mu_i$ with $\mu =
%1,2,3$. Each of the spin glass phases ``breaks" exactly two of
%these symmetries. %For example, when $J_z \gg J_x \sim J_y$,
%$\Pi^x$ and $\Pi^y$ are spontaneously ``broken"; $\Pi^z$ generates
%a residual $\dsZ_2$ symmetry; and so the spin glass has a
%$\dsZ_2^2 / \dsZ_2 = \dsZ_2$ or two-fold degenerate spectrum.

Beside the exact global symmetry $\dsZ_2^2$, the model also has
some statistical symmetries, which are valid only in the
statistical sense over the ensemble of the disordered
Hamiltonians. When $J_x = J_y$ (and similarly for $J_y = J_z$ and
$J_z = J_x$), the \emph{distribution} of Hamiltonians $H$ has a
$\dsZ_2$ symmetry which swaps $J_{i,x} \leftrightarrow J_{i,y}$.
%This $\dsZ_2$ symmetry allows us to fine tune our tuning parameters into the critical phase.
When $J_x = J_y = J_z$, the distribution of Hamiltonians has an
$S_3$ permutation symmetry which permutes $J_{i,x} \leftrightarrow
J_{i,y} \leftrightarrow J_{i,z}$.
For any $J_\mu$, the distribution of Hamiltonians also has %$\dsZ_L$
translation symmetry. Imposing these statistical symmetries can be
used to easily fine tune the XYZ model to its critical phases.

In the existing literature, the RSRG-X approach
\cite{Altman:2014hg,Potter:2015th,Refael:2013du,Vasseur:2015ys}
has been successfully applied to analyze various marginal MBL
phases in disordered spin systems. However, it is challenging to
apply the traditional closed-form RSRG-X analysis to the XYZ
model, even near the free fermion soluble points (such as
$J_x=J_y$ and $J_z\to 0$). At the free fermion soluble point, the
spin system can be mapped to two independent Majorana chains with
uncorrelated randomness, which allows the standard bond decimation
RG scheme to be applied independently on each chain. However, once
the fermion interactions ($J_z$ terms) are introduced to the
system, the two Majorana chains are coupled together as a ladder
lattice. The independent bond decimation on both chains will
quickly distort the ladder lattice and generate complicated
configurations of multi-fermion interactions, which can not be
tracked in closed-form. Therefore, we turn to the numerical
approach of SBRG, which can keep track of all orders of
multi-fermion interactions generated under the RG flow.

In the following, we will study the XYZ model by applying SBRG. We
will show results for $\Gamma = 4$ initial randomness,
%because this is the smallest $\Gamma$
for which SBRG agrees well with exact diagonalization on small
latices (data not shown in this paper), and
%for which
our approximations appear to be safe (see \appref{sec:SBRG
Approximations} and \ref{sec:ED}, and \secref{sec: MBL vs
thermal}). We will also limit our system sizes to $L \leq 256$
because our current implementation of SBRG does not produce
accurate results for larger system sizes on the critical lines of
the XYZ model.

\section{Phase Diagram}

\begin{figure}[thbp]
\includegraphics[width=\columnwidth]{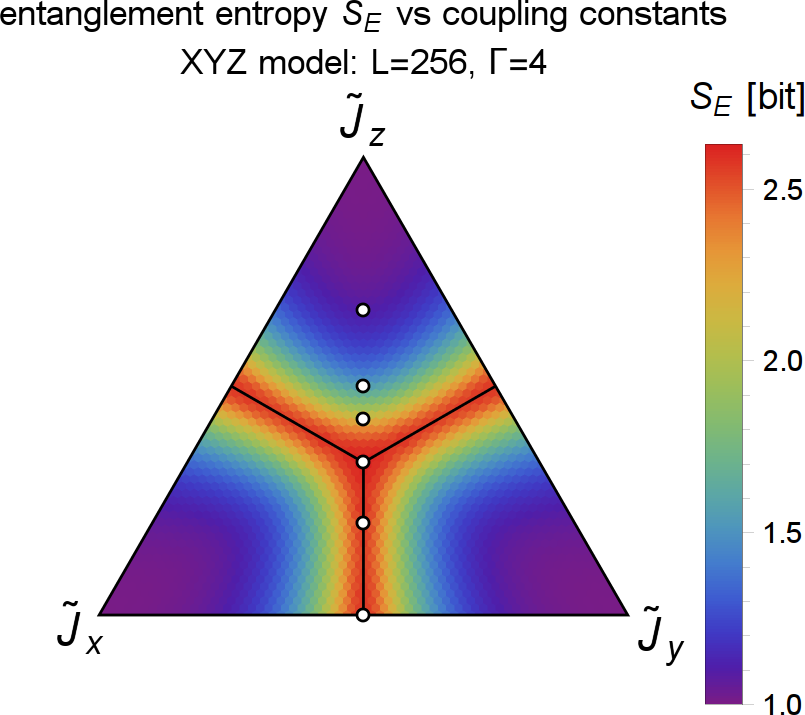}
\includegraphics[width=\columnwidth]{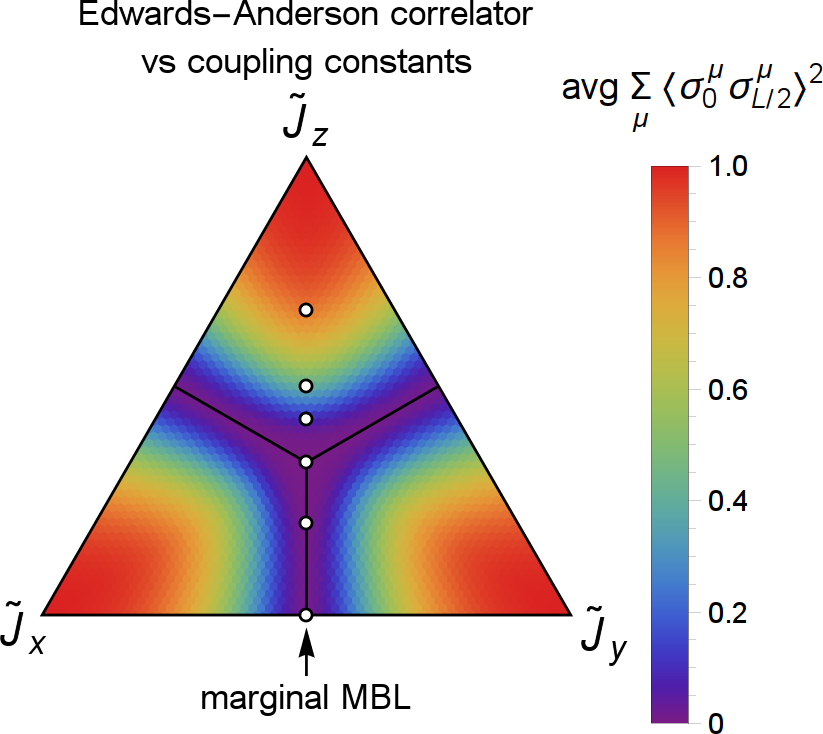}
\caption{ Ternary plot of the disorder and energy averaged
entanglement entropy $\SE$ [bit] ($\equiv \SE / \ln2$)
(\eqnref{eq:S_E}) of a subsystem of length $L/2$ (top) and
Edwards-Anderson correlator (\eqnref{eq:anderson}) (bottom) vs
coupling constants ($0 < \J_{x,y,z} < 1$) for the XYZ spin chain
of length $L=256$. We use this plot to sketch the phase
diagram. When $\J_z > \max(\J_x, \J_y)$, the system is in a
$\dsZ_2$ spin glass state. When $\J_z < \J_x = \J_y$, the system
is in a marginal MBL phase.
(The other phases are given by permutations of $x,y,z$.)
The white dots correspond to the
points in the phase diagram that are shown in
\figref{fig:entanglement},~\ref{fig:anderson},~\ref{fig:MI}.
}\label{fig:phase diagram}
\end{figure}

\subsection{Spin Glass Phases}

With large disorder and at finite energy density, there are three
spin glass phases (\figref{fig:phase diagram}). We find that if
$\J_z$ is the largest coupling constant (i.e. $\J_z > \J_x$ and
$\J_z > \J_y$), then the system is in an MBL spin glass phase
where the correlator $\sigma_i^z \sigma_j^z$ shows long range
glassy behavior. That is, $\sigma_i^z \sigma_j^z$ develops a
finite overlap with products of the local integrals of motion
(LIOM) $\tau^z$s of the MBL phase (see \eqnref{eq:H_eff} in
\appref{sec: MBL}), and is thus roughly conserved. This finite
overlap results in an Edwards-Anderson correlator that asymptotes
to a nonzero constant at large distance, which we take to be the
primary signature of a spin glass phase.

\subsection{Critical Phases}

With sufficiently strong disorder, the spin glass phases appear to
be separated by a critical lines (e.g. $\J_z \leq \J_x = \J_y$)
consisting of marginal MBL phases, which is evidenced by the fact
that the entanglement entropy diverges logarithmically
(\figref{fig:entanglement}) in this phase. In \figref{fig:anderson
Jz} we provide evidence that the critical ($\J_z \leq \J_x =
\J_y$) to spin glass ($\J_z > \J_x = \J_y$) phase transition is
continuous and occurs exactly at $J_x = J_y = J_z$ by showing
evidence that the long-range spin glass Edwards-Anderson
correlator is zero when $\J_z \leq \J_x = \J_y$ and increases
continuously for $\J_z > \J_x = \J_y$. An example of how the LIOM
look in this phase is shown in \figref{fig:holography critical}.

\begin{figure}[thbp]
\includegraphics[width=\columnwidth]{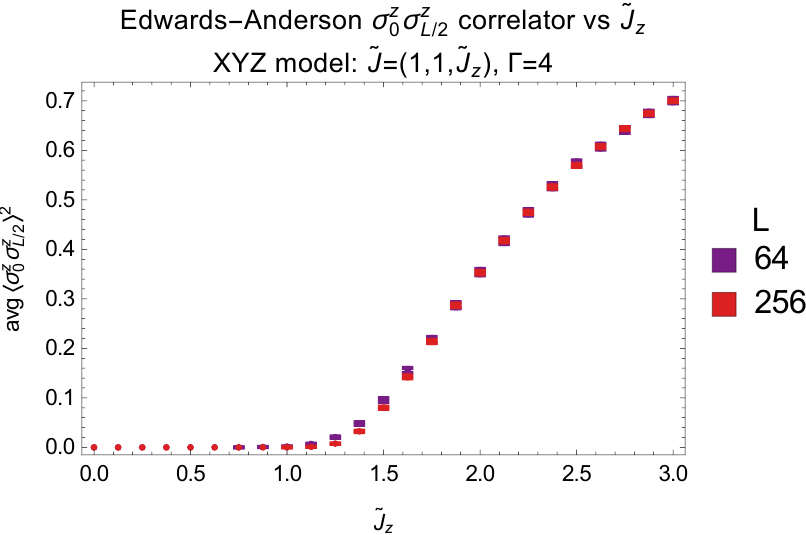}
\caption{ Disorder and energy averaged Edwards-Anderson correlator
(\eqnref{eq:anderson}) at separation $x=L/2$ vs $\J_z$ for
different system sizes $L$ in the XYZ spin chain with $\J_x = \J_y
= 1$. In the critical phase, the Edwards-Anderson correlator
decays algebraically to zero; while in the spin glass phase, the
Edwards-Anderson correlator asymptotes to a constant. Furthermore,
as one tunes $\J_z$ from the critical phase to the spin glass
phase, the Edwards-Anderson correlator becomes nonzero
continuously at $\J_z = 1$, indicating that this marks the
location of a continuous phase transition. \footError
}\label{fig:anderson Jz}
\end{figure}

\subsubsection{Marginal MBL vs Thermal}
\label{sec: MBL vs thermal}

With sufficiently large disorder, SBRG depicts the critical phase
as marginally MBL as SBRG finds a logarithmically diverging
entanglement entropy in this phase. However, SBRG is not capable
of describing thermalization, and so one should worry that
resonance effects may thermalize the critical phase. Indeed, the
instability of marginal MBL phase to thermalization has been
demonstrated in other 1D spin models.
\cite{Potter:2015th,Vasseur:2015ys} Thus, it is crucial to check
the approximations made by SBRG, in order to verify that an exact
RG does not flow toward thermalization.
In \appref{sec:ED} we do a brief exact diagonalization study to check that with strong disorder, the system is not thermal.
Below, we will study the evidence against thermalization by using SBRG.

The core approximation
made by SBRG is the validity of the second order perturbation
theory used by each RG step. This and other approximations are
explained in detail and accounted for in \appref{sec:SBRG
Approximations}. However, even though these approximations are
controlled by strong disorder, one could still worry that errors
may build up during the RG flow and cause SBRG to incorrectly
depict the critical phase as a marginal MBL phase, even if the
critical phase is actually thermal.

At intermediate RG steps, a cluster of $n$ ``LIOM'' could resonate
and become thermalized by an off-diagonal term that mixes (i.e.
anticommutes with) them if the energy of the off-diagonal term
$\varepsilon_\text{mix}$ is larger than the smallest energy
difference $\Delta E$ of the $n$ ``LIOM'': i.e. if
$\varepsilon_\text{mix} > \Delta E$. The verified assumptions
\eqnref{eq:wolffError} and \ref{eq:offdiagError} imply that this
rarely occurs for small $n$. However, one might worry that large
clusters of $n$ ``LIOM'' could be thermalized by rare off-diagonal
terms \cite{Huse:2015cd}. But a cluster of $n$ LIOM will describe
$2^n$ states, and thus typically have a smallest energy level
spacing equal to  $\Delta E \sim 2^{-n}$ (because our model has a
small symmetry, there is no symmetry protected degeneracy in this
cluster); while below we argue that off-diagonal terms at
intermediate RG steps will have energies of order
$\varepsilon_\text{mix} \lesssim e^{-\Gamma n}$, which is much
smaller than $\Delta E \sim 2^{-n}$ for large disorder $\Gamma$.
Thus, $\varepsilon_\text{mix} \ll \Delta E$ and so it seems
unlikely that enough clusters of $n$ ``LIOM'' could resonate to
thermalize the system.

\begin{figure}[thbp]
\includegraphics[width=\columnwidth]{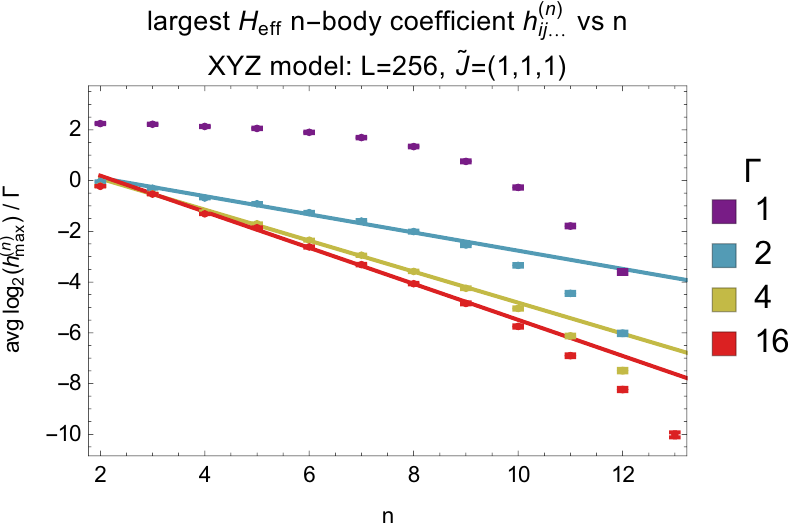}
\caption{ Disorder average of $\log_2(h_\text{max}^{(n)}) /
\Gamma$ vs $n$ for different amounts of disorder $\Gamma$ where
$h_\text{max}^{(n)} \equiv \max_{ij\dots} |h_{ij\dots}^{(n)}|$
(\eqnref{eq:h_max}) is the largest n-body coefficient
$h_\text{max}^{(n)}$ of the effective Hamiltonian $H_\text{eff}$
at the critical point $\J_x = \J_y = \J_z = 1$. As the disorder
$\Gamma$ increases, the data converges to a single straight line.
For large $n$, the data drops below the linear fit line. This is
expected to be an artifact of dropping terms in SBRG (see
\secref{sec:SBRG details} for details). Solving for
$h_\text{max}^{(n)}$ gives $h_\text{max}^{(n)} \sim e^{-\Gamma n}$
as in \eqnref{eq:h_max}, which completes our argument that the
critical line is marginally MBL. \footError }\label{fig:nBody}
\end{figure}

To show that $\varepsilon_\text{mix} \lesssim e^{-\Gamma n}$, we
note that at intermediate RG steps, the energies of off-diagonal
terms mixing $n$ ``LIOM'' should be roughly bounded by the the
largest n-body coefficient $h_\text{max}^{(n)}$ of the effective
Hamiltonian $H_\text{eff}$ (\eqnref{eq:H_eff}):
$\varepsilon_\text{mix} \lesssim h_\text{max}^{(n)}$. And in
\figref{fig:nBody} we show that $h_\text{max}^{(n)}$ decays
exponentially with $\Gamma\, n$:
\begin{align}
\label{eq:h_max}
h_\text{max}^{(n)} &\equiv  \max_{ij\dots} |h_{ij\dots}^{(n)}| \\
                   &\sim    e^{-\Gamma n} \nonumber
\end{align}
Thus $\varepsilon_\text{mix} \lesssim h_\text{max}^{(n)} \sim e^{-\Gamma n}$.

\section{Entanglement Entropy}
\label{sec:S_E}

The entanglement entropy is a useful tool to probe quantum phase
transitions from the entanglement patterns of the many-body state,
and has the nice property that one does not have to pick the right
order parameter. Instead, one only has to choose a useful
subsystem geometry. SBRG can efficiently calculate
\cite{You:2015sb} entanglement entropies using the stabilizer rank
algorithm introduced by \refcite{Chuang:2004ni}.

The entanglement entropy $\SE$ of a subsystem $A$ for a
wavefunction $\ket{\psi}$ is defined to be
\begin{align}
  \SE(A) &\equiv - \Tr[\rho_A \ln \rho_A] \label{eq:S_E}\\
  \rho_A &=        \Tr_{\overline A} \; \ket{\psi}\bra{\psi} \nonumber
\end{align}
where $\Tr_{\overline A}$ means that degrees of freedom not in $A$
are traced out. The disorder configuration ($\delta$) and energy
($E$) averaged entanglement entropy is then
\begin{align}
\text{avg } \SE(A) &\equiv \frac{1}{N_\delta} \sum_\delta
\frac{1}{N_E} \sum_E \SE(A)
\end{align}
where $\psi$ and thus $\rho_A$ in \eqnref{eq:S_E} depend on
$\delta$ and $E$. We average over all energy eigenstates because
we're interested in the entire spectrum of states. Additionally,
in the strong disorder limit, the LIOM take the form of products
of Pauli matrices, and the eigenstates of these LIOM (and the
Hamiltonian) all have the same entanglement entropies. However, this is only a special feature of Pauli-like LIOM in the strong disorder limit. Away from the strong disorder limit, the LIOM will be further dressed by higher order corrections, such that different eigenstates will not have identical entanglement structures. Nevertheless, the difference of entanglement entropies across the spectrum will be relatively small in the strong disorder regime. So we will neglect the spectral dependence, and consider the energy averaged entanglement entropy.

 First, we will take the subsystem $A$ to be a line segment of length $\ell$.
We will be interested in how the entanglement entropy scales as
the subsystem length $\ell$ increases. Later, in \secref{sec:LRMI}
we will use more complicated subsystem geometries in order to
study the long range mutual information of the XYZ model.
\begin{figure}[thbp]
\includegraphics[width=\columnwidth]{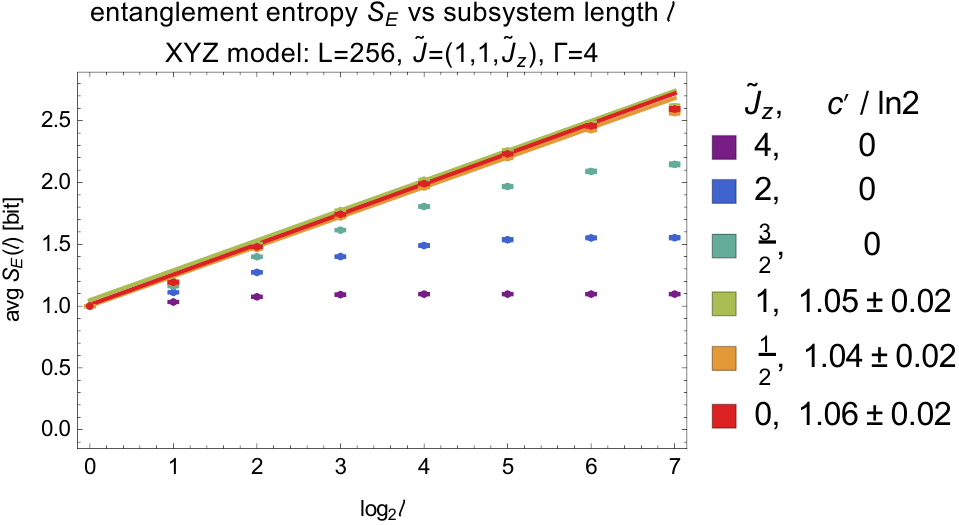}
\includegraphics[width=\columnwidth]{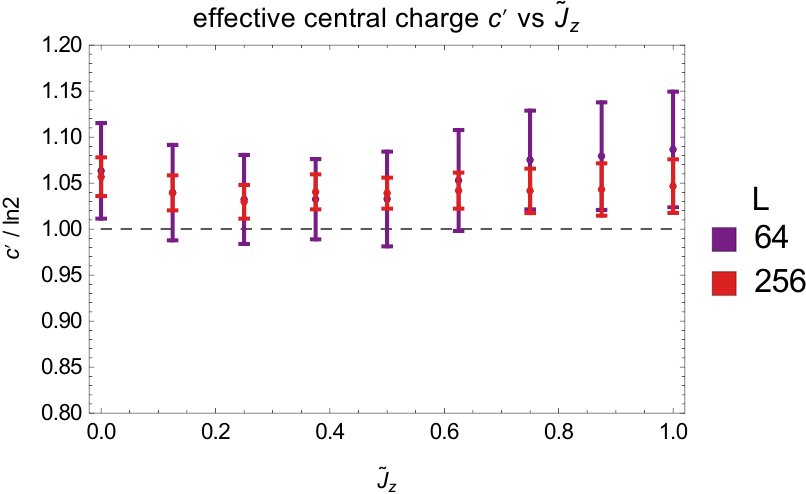}
\caption{ \textbf{(top)} Disorder and energy averaged entanglement
entropy $\SE$ [bit] ($\equiv \SE / \ln2$) (\eqnref{eq:S_E}) vs
subsystem length $\ell$ for different $\J_z$ in the XYZ spin chain
with $\J_x = \J_y = 1$ and system size $L=256$. When $\J_z > 1$,
the system is a spin glass and the entanglement entropy follows an
area law for large subsystems. When $\J_z$ is only slightly larger
than 1, e.g. $\J_z = 3/2$, there appears to be large finite size
effects. When $\J_z \leq 1$, the system is critical and the
entanglement entropy follows a log law $\SE(\ell) \sim
\frac{c'}{3} \ln\ell$, and the effective central charge $c'$ is
estimated using the slope of the fit to the $\ell=\frac{1}{2}\sqrt{L},\sqrt{L},2\sqrt{L}$ data points.
\textbf{(bottom)} Effective central charge $c'$
(\eqnref{eq:central charge}) vs $\J_z$ for different system sizes
$L$ in the XYZ spin chain with $\J_x = \J_y = 1$. For all $\J_z <
1$, the effective central charge appears to be consistent with $c'
= c\ln2$ \cite{Moore:2004ee} where $c=1$ is the central charge
without disorder. \footError }\label{fig:entanglement}
\end{figure}
As shown in \figref{fig:entanglement}, in the spin glass phases,
the entanglement entropy $\SE$ asymptotes to a constant with
increasing subsystem length $\ell$ since it is a short range
correlated phase. Furthermore, $\SE \geq \ln2$ due to the spin
glass order. These results can be understood deep in the spin
glass phase via the LIOM in \eqnref{eq:SG LIOM} from the
stabilizer rank algorithm. \cite{Chuang:2004ni, You:2015sb} See
\appref{sec:SG EE} for details.

% \subsubsection{Critical Phases}

In the critical phases, the entanglement entropy $\SE$ appears to
diverge logarithmically with subsystem length $\ell$ (\figref{fig:entanglement}):
\begin{align}
  \SE(\ell) &= \frac{c'}{3} \ln\ell \label{eq:central charge}\\
         c' &= \ln2                 \nonumber
\end{align}
The constant $c'$ is the effective central charge, which is
postulated to be related to the central charge $c$ without
disorder by $c' = c \ln2$ \cite{Moore:2004ee} (for Ising and
Heisenberg types of models). Without disorder, the XYZ model has
$c=1$ in the critical phase, and so we expect and observe $c' =
\ln2$ in our disordered model (\figref{fig:entanglement}). The
logarithmically diverging entanglement entropy in
\eqnref{eq:central charge} is a result of the LIOM becoming
nonlocal (see \figref{fig:holography critical}). This causes more LIOM to be cut by the subsystem $A$ and
enter the entanglement entropy equation \eqnref{eq:H_eff EE}.

\section{Edwards-Anderson Correlator}
\label{sec:anderson}

A common approach to characterize glassy order is the
Edwards-Anderson correlator. The disorder and energy averaged
Edwards-Anderson spin glass correlator of $\sigma_0^\mu
\sigma_x^\mu$ is defined to be
\begin{equation}
\label{eq:anderson} \text{avg} \braket{\sigma_0^\mu
\sigma_x^\mu}^2 \equiv \frac{1}{N_\delta} \sum_\delta
\frac{1}{N_E} \sum_E \braket{E | \sigma_0^\mu \sigma_x^\mu |
E}^2_\delta
\end{equation}
It is the average of the square of $\braket{\sigma_0^\mu
\sigma_x^\mu}$ over disorder configurations ($\delta$) and
energies ($E$). $\sigma_0^\mu$ and $\sigma_x^\mu$ are Pauli
matrices at lattice sites $0$ and $x$, respectively. We average
over all energy eigenstates because we're interested in the entire
spectrum of states. Additionally, in the strong disorder limit,
the LIOM take the form of products of Pauli matrices, and the
eigenstates of these LIOM (and the Hamiltonian) all have the same
Edwards-Anderson Correlators.

\begin{figure}[thbp]
\includegraphics[width=\columnwidth]{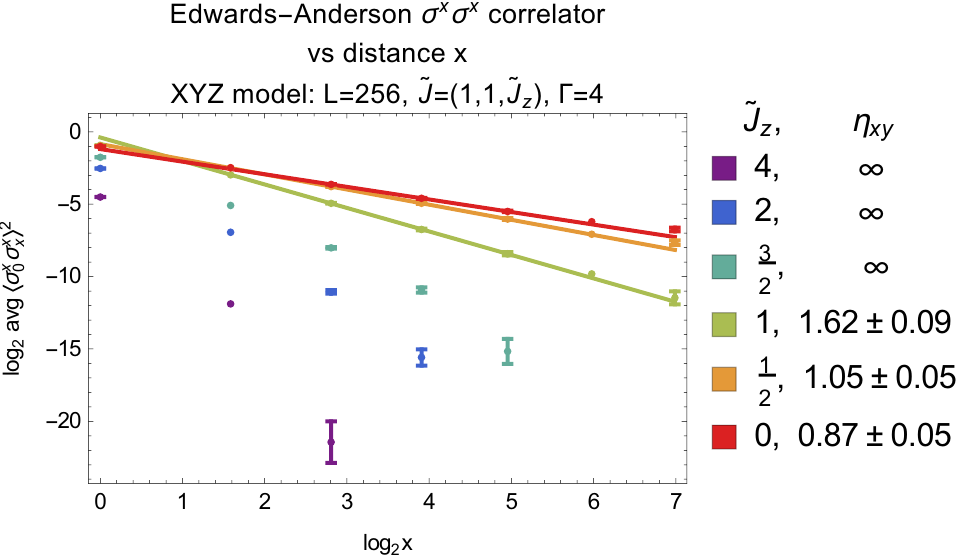}
\includegraphics[width=\columnwidth]{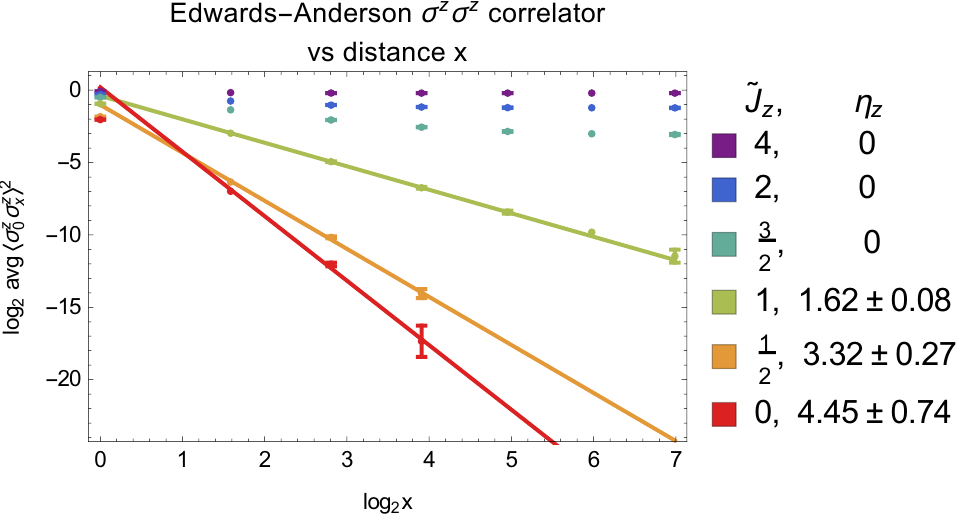}
\caption{ Disorder and energy averaged Edwards-Anderson
$\sigma^x\sigma^x$ (top) and $\sigma^z\sigma^z$ (bottom)
correlators (\eqnref{eq:anderson}) vs distance $x$ for different
$\J_z$ in the XYZ spin chain with $\J_x = \J_y = 1$ and system
size $L=256$. When $\J_z > 1$, the system is a $\sigma^z \sigma^z$
spin glass phase, and the $\sigma^x \sigma^x$ correlator decays
exponentially while the $\sigma^z \sigma^z$ correlator asymptotes
to a constant.
% The exponential decay is not clear from the figure due to a very large correlation length.
When $\J_z$ is only slightly larger than 1, e.g.
$\J_z = 3/2$, there appears to be moderate finite size effects.
When $\J_z < 1$, the system is critical with power law $\sigma^\mu
\sigma^\mu$ correlators ($\mu$ summation not implied) with critical exponents $\eta_{xy}$ and $\eta_z$
(\eqnref{eq:eta}, \figref{fig:eta}), which were calculated using the
slope of the fit to the $x=\frac{1}{2}\sqrt{L}-1,\sqrt{L}-1,2\sqrt{L}-1$ data points.
% Some data points are missing for large distances because not enough terms in $\tilde Q$ were used to obtain accurate date.
\footError }\label{fig:anderson}
\end{figure}

\begin{figure}[thbp]
\includegraphics[width=\columnwidth]{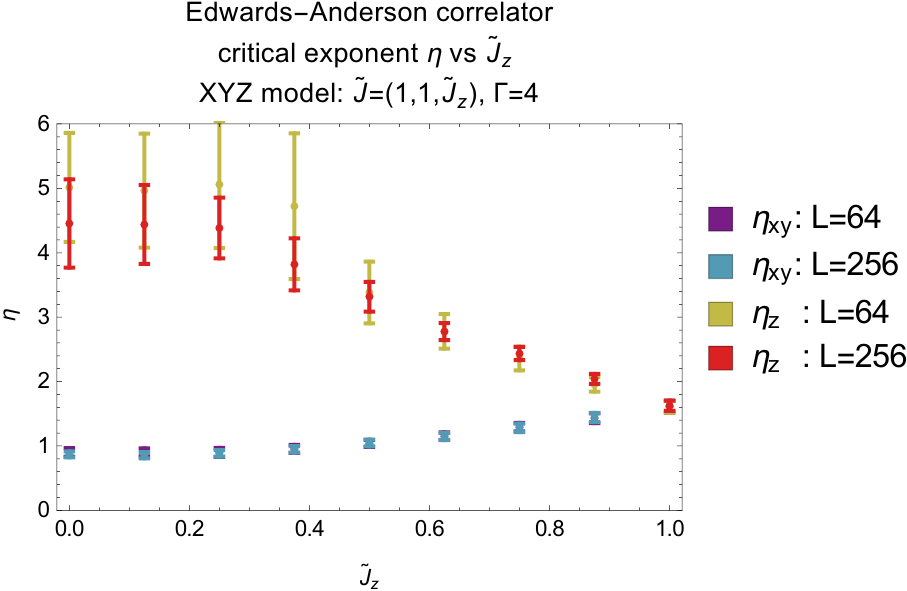}
\caption{ Critical exponents $\eta_{xy}$ and $\eta_z$
(\eqnref{eq:eta}) vs $\J_z$ with $\J_x = \J_y = 1$ for the XYZ
spin chain of various system sizes $L$. $\eta_{xy}$ ($\eta_z$)
appears to increase (decrease) continuously as $\J_z$ increases.
At least for $\J_z > 1/2$, we do not expect the continuously varying exponents
to be a result of finite size effects because $\eta$ shows very little system size dependence in the above plot.
The critical exponents were calculated as shown in
\figref{fig:anderson}. \footError }\label{fig:eta}
\end{figure}

% \subsubsection{Spin Glass Phases}

If $\J_z > \J_x$ and $\J_z > \J_y$, the system is in a $\sigma^z
\sigma^z$ spin glass phase and the $\sigma^z \sigma^z$
Edwards-Anderson correlator (\eqnref{eq:anderson}) asymptotes to a
constant for large distance $x$ (with exponentially small
corrections) (\figref{fig:anderson}). Physically, this implies
that $\sigma_i^z \sigma_j^z$ has developed a finite overlap with
products of the local integrals of motion (LIOM)
(\eqnref{eq:H_eff}) of the MBL phase, and is thus roughly
conserved. However, the $\sigma^x \sigma^x$ and $\sigma^y
\sigma^y$ correlators decay exponentially with distance $x$, as
expected.

% \subsubsection{Critical Phases}

When $\J_z < \J_x = \J_y$, the system is in a marginal MBL
critical phase between $\sigma^x \sigma^x$ and $\sigma^y \sigma^y$
spin glasses (\figref{fig:phase diagram}). The spin configuration
in the marginal MBL state is dominated by nested domains of
$\sigma^x\sigma^x$ and $\sigma^y\sigma^y$ Ising spin glass orders. The domains have fractal
structures throughout the lattice, leading to the power-law decay
of $\sigma^\mu \sigma^\mu$ Edwards-Anderson
correlators ($\mu$ summation not implied) with critical exponents $\eta_{xy}$ and $\eta_z$
(\figref{fig:eta}):
\begin{align}
\text{avg} \braket{\sigma_0^x \sigma_x^x}^2 &=
\text{avg} \braket{\sigma_0^y \sigma_x^y}^2 \sim x^{-\eta_{xy}} \nonumber\\
\text{avg} \braket{\sigma_0^z \sigma_x^z}^2 &\sim x^{-\eta_z} \label{eq:eta}
\end{align}
Unlike the effective central charge $c'$ (\eqnref{eq:central
charge}) which remains fixed, $\eta_{xy}$ ($\eta_z$) appears to increase (decrease)
continuously with increasing $\J_z$.
The power-law decay of $\sigma^z \sigma^z$ can be understood in
% The $\sigma^z \sigma^z$ Edwards-Anderson correlator also decays as
% a power-law when $\J_z < \J_x = \J_y$, which becomes obvious in
the limit $J_z = 0$, where the system can be mapped to two
independent free random Majorana fermion chains.

\section{Long Range Mutual Information}
\label{sec:LRMI}

\begin{figure}[thbp]
\includegraphics[width=\columnwidth]{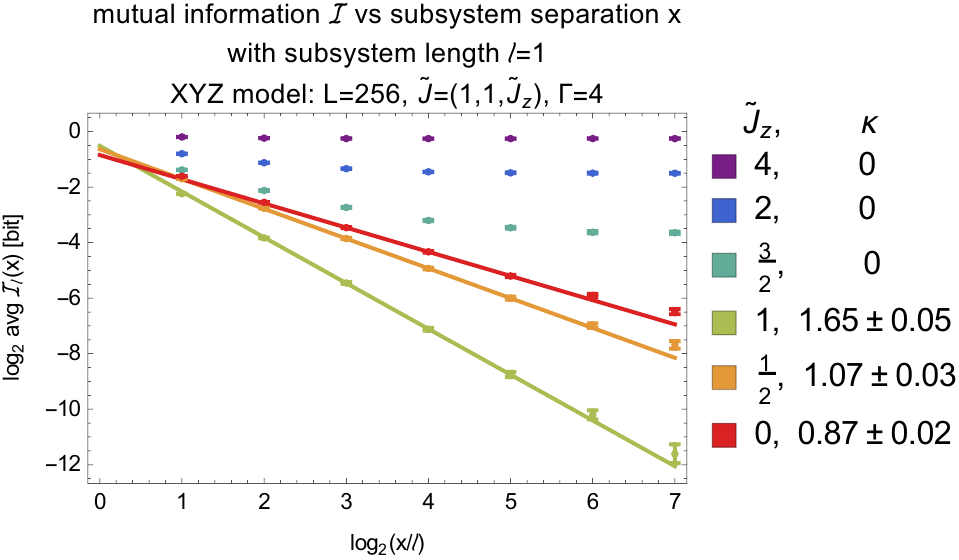}
\includegraphics[width=\columnwidth]{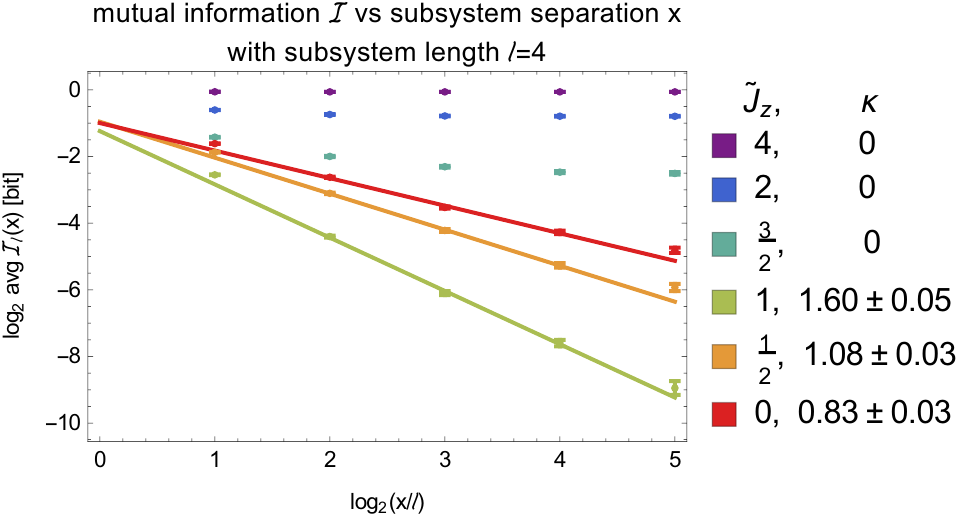}
\includegraphics[width=\columnwidth]{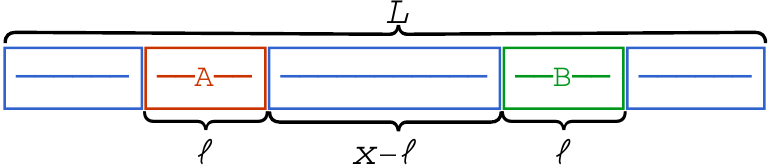}
\caption{ Disorder and energy averaged long range mutual
information (LRMI) $\mathcal{I}$ of two subsystems (of length
$\ell = 1$ (top) and $\ell = 4$ (middle)) vs subsystem separation
$x$ for different $\J_z$ in the XYZ spin chain with $\J_x = \J_y =
1$ and system size $L=256$. Subsystem length $\ell$ and separation
$x$ are defined as shown in the bottom diagram. When $\J_z > 1$,
the system is in a spin glass phase, and the LRMI asymptotes to a
constant.
When $\J_z$ is only slightly larger than 1, e.g.
$\J_z = 3/2$, there appears to be moderate finite size effects.
When $\J_z \leq 1$, the system is in a marginal MBL
phase, and the LRMI decays according to a power law with a
critical exponent $\kappa$: $\mathcal{I}_\ell(x) \sim
(x/\ell)^{-\kappa}$. The critical exponent $\kappa$ does not
appear to depend on the subsystem length $\ell$, but it does
appear to increase continuously with $\J_z$, unlike the effective
central charge $c'$ which remains fixed
(\figref{fig:entanglement}). If the above plots were drawn on top
of each other, the critical $\J_z \leq 1$ data would very nearly
overlap, which implies that changing $\ell$ just rescales $x$ as
one would expect in a scale invariant system:
$\mathcal{I}_\ell(x/\ell) \approx \mathcal{I}_{\ell'}(x/\ell')$.
\footError }\label{fig:MI}
\end{figure}

In \secref{sec:S_E} we used the entanglement entropy to diagnose
the critical phase as a marginal MBL phase due to its
logarithmically diverging entanglement entropy (\eqnref{eq:central
charge}). However, only studying the asymptotics of the
entanglement entropy $\SE(\ell)$ for large connected subsystem has
some limitations. For example, $\SE(\ell)$ can not firmly
distinguish a trivial phase from a spin glass, since both just
asymptote to a constant. Instead of studying the entanglement
entropy of just a single connected subsystem, it has been proposed
to study the long range mutual information (LRMI) between two
disconnected subsystems with large separation \cite{Qi:2015mi}.
The mutual information $\mathcal{I}$ between two non-overlapping
subsystems $A$ and $B$ is defined to be
\begin{align}
  \mathcal{I}(A,B) = \SE(A) + \SE(B) - \SE(A \cup B)
\end{align}
We will take subsystems $A$ and $B$ to be lines of length $\ell$
with separation given by $x$ as shown in \figref{fig:MI} (bottom).
We will be interested in the $L=\infty$ limit of the asymptotics
of the LRMI $\mathcal{I}_\ell(x)$ as the separation $x$ tends to
infinity while the subsystem length $\ell$ is held fixed. Since the mutual information is also an upper bound of correlation functions, in principle, the mutual information should decay slower than any correlation function.

In a direct product state with no spin glass order (which doesn't
exist in the XYZ model), far apart regions share little
entanglement, and thus the LRMI should decay exponentially. In a
spin glass phase there is long range glassy order, and the LRMI
asymptotes to a constant (\figref{fig:MI}). Thus, LRMI can easily
distinguish a trival phase from a spin glass phase. This fact
could be very useful in more complicated models where finding the
right order parameter is difficult. Finally, in a critical phase
in any dimension, the LRMI is expected to decay according to a
power law with some critical exponent $\kappa$:
$\mathcal{I}_\ell(x) \sim (x/\ell)^{-\kappa}$. In the XYZ model, a
power law is indeed observed (\figref{fig:MI}). Since the mutual information should not decay faster than the correlation functions,  we at least expect $\kappa\leq \min(\eta_{xy},\eta_z)$, which is also verified within the numerical error. The critical
exponent $\kappa$ does not seem to depend on the subsystem length
$\ell$; but $\kappa$ does appear to increase continuously with
$\J_z$, unlike the effective central charge $c'$ which remains
fixed. Thus, the LRMI can efficiently tell if a phase is a
trivial, spin glass, or marginal MBL critical phase; and works as expected in
the XYZ model (\figref{fig:MI}).

\section{Conclusion}

In conclusion, we have studied the XYZ spin chain with independently
random XX, YY, and ZZ couplings on each bond. Unlike the random XXZ
or Heisenberg models, the XYZ model breaks the continuous spin
rotational symmetry down to the discrete $\dsZ_2^2$ symmetry, such
that the quantum phase transitions between different
symmetry-breaking spin glass phases can persist to finite energy
density as marginal MBL critical lines. We use the SBRG numerical
method to calculate the entanglement entropy, Edwards-Anderson
correlator, and long-range mutual information. In the MBL spin
glass phase, the entanglement entropy follows the area-law scaling
and quickly saturates to a value of $\SE\geq\ln 2$. Both the
Edwards-Anderson correlator and the long-range mutual information
exhibit long-range behavior, demonstrating spin glass
order. Along the marginal MBL critical line, the entanglement
entropy follows the logarithmic scaling $\SE(\ell)=(c'/3) \ln
\ell$, with a fixed effective central charge $c'=\ln 2$. Both the
Edwards-Anderson correlator and the long-range mutual information
decays in a power-law, and the critical exponents varies
continuously along the marginal MBL line.

\acknowledgments
We acknowledge helpful discussions with David Huse and Andrew Potter.
In particular, \appref{sec:ED} resulted from comments from David Huse.
The authors are supported by David and Lucile Packard foundation and NSF Grant No. DMR-1151208.

\bibliography{XYZSBRG}

% \onecolumngrid
% \twocolumngrid
\newpage
\appendix

\section{Many Body Localization}
\label{sec: MBL}
A Hamiltonian is said to be fully many body localized (MBL) if all eigenstates in the many-body spectrum are localized \cite{Huse:2015rv}.
%(In this work,
%we will avoid Hamiltonians where only some of the eigenstates are
%MBL.) Currently, all known MBL Hamiltonians result from strongly
%breaking translational symmetry through disorder. MBL systems
%should be contrasted with most translationally invariant
%Hamiltonians, which do serve as their own heat bath, and are
%therefore said to be thermal.
%Thus, unlike a thermal system, even if a heat bath is placed next
%to the boundary of a fully MBL system, the system will not
%thermalize. Instead, thermalizing a fully MBL system requires
%connecting a heat bath to the bulk of the system.
It is believed that there exists a finite local unitary
transformation $U$ that can diagonalize the MBL Hamiltonian $H$:
\begin{align}
\label{eq:H_eff}
  H_\text{eff} &= U H U^\dagger \\
               &= \sum_i h_i^{(1)} \tau_i^z + \sum_{ij} h_{ij}^{(2)} \tau_i^z \tau_j^z + \sum_{ijk} h_{ijk}^{(3)} \tau_i^z \tau_j^z \tau_k^z + \cdots \nonumber
\end{align}
where the $\tau_i^z$ are the LIOM. The fact that $U$ is a finite
local unitary transformation means that it can be written as a
finite time evolution of a time dependent local Hamiltonian with
bounded spectrum. This implies that the LIOM $U \tau_i^z
U^\dagger$ must be local operators (with exponentially decaying
tails). This implies that the eigenstates of an MBL Hamiltonian
display an area law entanglement (\figref{fig:entanglement}), as
opposed to the volume law seen in excited states of thermal
systems. An example of such a $U$ can be inferred from
\figref{fig:holography SG}. Furthermore, the $h_{ijk\dots}^{(n)}$
decay exponentially with $n$ (\figref{fig:nBody}) and distance
$\max(|i-j|, |i-k|, |j-k|, \dots)$. This implies that in an MBL
system, the time evolution of a direct product state displays an
entanglement entropy which increases only logarithmically with
time \cite{Abanin:2013ta,Moore:2012ge,Znidaric:2008he}, instead of
linearly as in thermal systems. This implies that an MBL system
can't efficiently spread entanglement, and thus can't act as its
own heat bath.

\subsection{Marginal MBL}

If one observes a phase transition between two fully MBL phases,
then it's possible that the critical point between them is a
marginal MBL phase. A margin MBL phase doesn't obey
\eqnref{eq:H_eff} with the same restrictions for the unitary
transformation $U$ and coefficients $h_{ijk\dots}^{(n)}$. For
example, the LIOM in the original bases $U \tau_i^z U^\dagger$
become nonlocal \cite{You:2015sb}, which implies that $U$ can't be
a finite local unitary transformation. In SBRG, the unitary
transformation $U$ that is found is nonlocal and requires a time
evolution that diverges logarithmically with the system size, as
can be inferred from \figref{fig:holography critical}.

\section{Spectrum Bifurcation Renormalization Group}
\label{sec: SBRG} In this work, we use the recently developed
Spectrum Bifurcation Renormalization Group (SBRG)
\cite{You:2015sb} to simulate the XYZ model. SBRG is similar to
RSRG-X \cite{Altman:2014hg}, and behaves similarly to RSRG-X for
the models on which RSRG-X can be applied. However, SBRG differs
in that it (approximately) computes the commuting local integrals
of motion (LIOM) (also know as localized-bit \cite{Huse:2014ec}
stabilizers) of a Hamiltonian, and therefore targets the entire
spectrum at once; while RSRG-X targets a specific eigenstate
energy at a time. That is, given an arbitrary local Hamiltonian
$H$ written in terms of physical spins $\sigma_i^\mu$, SBRG
computes the unitary transformation $U$ to a set of LIOM Pauli
operators $\tau_i^\mu$ such that $H$ can be written as the sum of
products of commuting $\tau_i^z$ operators with coefficients
($h_i^{(1)}, h_{ij}^{(2)}, h_{ijk}^{(3)}, ...$) as in
\eqnref{eq:H_eff}. The unitary transformation $U$ is encoded using
alternating $C_4$ transformations (see end of \appref{sec:SBRG
Approximations}) and small Schrieffer-Wolff perturbations
(\eqnref{eq:wolff S}).

With this information, SBRG can efficiently compute many
quantities of interest such as Edwards-Anderson spin glass order
parameters, the energy spectrum, entanglement entropies, and other
LIOM properties. One advantage of SBRG is that it can handle
Hamiltonian terms which are arbitrary local products of sigma
matrices. This should be contrasted with other real space RG
schemes (such as RSRG-X \cite{Altman:2014hg}) which require a
specific Hamiltonian that is of closed form under RG. For example,
next nearest neighbor terms often are not allowed to be created by
the RG step of other methods. This flexibility allows SBRG to be
applied to a large class of spin systems in all dimensions,
including systems with topological order and symmetry protected
topological order.

However, the price paid for this generality is that the
Hamiltonian is not of closed form under RG, which results in
exponentially many terms generated by the RG flow. Many of these
terms will need to be dropped in order to preform computations
efficiently. In this work, up to 256 addition off-diagonal terms
are allows to be added during each RG step. This approximation
seems to work well for systems with large disorder and deep in the
fully MBL phase; but in marginal MBL phases, very large system
sizes can become problematic. However, for the XYZ chain, it
appears that reasonably accurate simulations can still be
performed for systems of roughly 256 spins in the marginal MBL
phase, and at least 10,000 spins deep in the MBL phase.

\subsection{RG Step and Approximations}
\label{sec:SBRG Approximations}

\begin{figure}[thbp]
\includegraphics[width=\columnwidth]{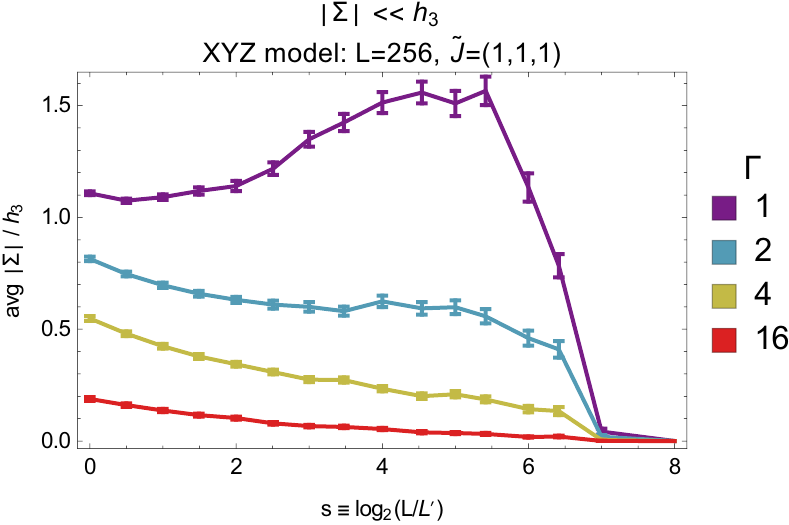}
\includegraphics[width=\columnwidth]{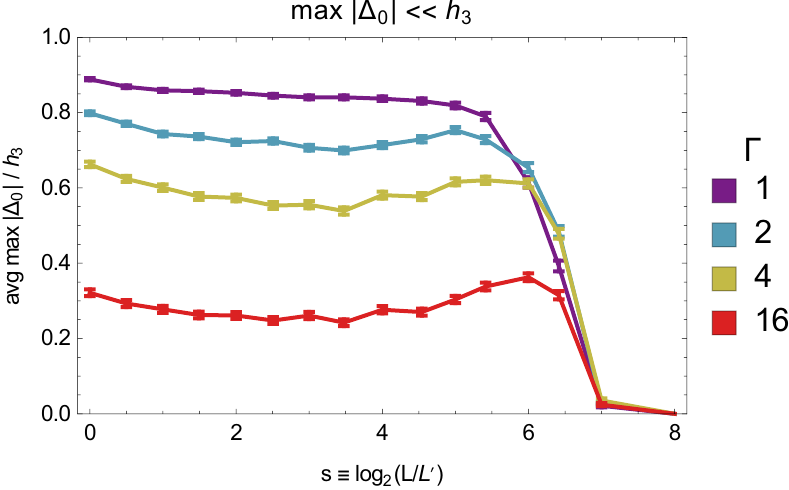}
\caption{ Disorder average of $|\Sigma| / h_3$
(\eqnref{eq:wolffError}) (top) and $\max |\Delta_0| / h_3$
(\eqnref{eq:offdiagError}) (bottom) vs amount of RG flow $s \equiv
\log_2(L/L')$ for different amounts of disorder $\Gamma$ where
$L=256$ is the system size and $L'$ is the number of remaining
spins. These are the small parameters used in the approximations
\eqnref{eq:wolffError} and \ref{eq:offdiagError}. Data is shown at
the critical point $\J_x = \J_y = \J_z = 1$; away from this point,
these approximations are even better. As the disorder $\Gamma$
increases, the plotted values get arbitrarily small, suggesting
that these approximations can be controlled by $\Gamma$. In
addition, the first approximation (top) gets better under RG flow
(for large $\Gamma$); the second (bottom) does not show a clear
trend.
Similar to \figref{fig:nBody}, the large $\Gamma$ data in this
plot can also be collapsed to a single curve if $f(y) =
y^{1/\Gamma}$ is applied to the data, which shows that $|\Sigma|
\sim \max |\Delta_0| \sim e^{-\Gamma}$ decreases exponentially as
$\Gamma$ increases. \footError }\label{fig:SBRG approx}
\end{figure}

In SBRG, the Hamiltonian of a system is written as a linear
combination of tensor products of pauli matrices:
\begin{equation}
  H = \sum_{[\mu]} h_{[\mu]} \sigma^{[\mu]} = \sum_{[\mu]} h_{[\mu]} \otimes_i \sigma^{\mu_i} \label{eq:initial H}
\end{equation}
As described in Appendix A2 of \cite{You:2015sb}, for each RG
step, the term $h_3 \sigma^{[\mu_3]}$ with the largest coefficient
$h_3$ is chosen to be the next local integral of motion (LIOM)
since it describes the leading energy scale. However,
$\sigma^{[\mu_3]}$ doesn't yet commute with the Hamiltonian $H$.
The Hamiltonian can be split into three parts
\begin{equation*}
  H = h_3 \sigma^{[\mu_3]} + \Delta + \Sigma
\end{equation*}
where $\Delta$ commutes with $\sigma^{[\mu_3]}$ while $\Sigma$
anticommutes with $\sigma^{[\mu_3]}$. In order to make
$\sigma^{[\mu_3]}$ into a LIOM, we must eliminate $\Sigma$. First
assume that
\begin{align}
  |\Sigma| &\ll h_3 \label{eq:wolffError}\\
  \max_{\Delta_0 \in \Delta, [\Sigma,\Delta_0] \neq 0} |\Delta_0| &\ll h_3 \label{eq:offdiagError}
\end{align}
where $|\Sigma| \equiv \sqrt{2^{-L} \,\Tr(\Sigma \cdot
\Sigma^\dagger)}$ and
  $\max_{\Delta_0 \in \Delta, [\Sigma,\Delta_0] \neq 0} |\Delta_0|$ is the absolute value of the largest pauli matrix coefficient of all terms in $\Delta$ that don't commute with $\Sigma$.
\figref{fig:SBRG approx} provides evidence that these assumptions are valid.
(RSRG-X depends on a similar set of assumptions.)
With these assumptions, a Schrieffer-Wolff
transformation can be performed in order to eliminate the
off-diagonal component $\Sigma$ to order $O(h_3^{-2})$:
\begin{align}
  S &= \exp\left( -\frac{1}{2h_3} \sigma^{[\mu_3]} \Sigma \right) \label{eq:wolff S}\\
    &= 1 - \frac{1}{2h_3} \sigma^{[\mu_3]} \Sigma - \frac{1}{8h_3^2} \Sigma^2 + O(h_3^{-3}) \nonumber\\
  H &\rightarrow H' = S^\dagger H S \nonumber\\
    &= h_3 \sigma^{[\mu_3]} + \Delta + \frac{1}{2h_3} \sigma^{[\mu_3]} \Sigma^2 \label{eq:H'}\\
    &+ \frac{1}{2h_3} \sigma^{[\mu_3]} [\Sigma, \Delta] + O(h_3^{-2}) \nonumber
\end{align}
The first approximation (\eqnref{eq:wolffError}) allows the
unitary $S$ to be expanded and implies that the third term in $H$
is small: $\frac{1}{2h_3} \sigma^{[\mu_3]} \Sigma^2 \sim
O(h_3^{-1})$. Although the first three terms of $H'$
(\eqnref{eq:H'}) commute with $\sigma^{[\mu_3]}$, the final term
($\frac{1}{2h_3} \sigma^{[\mu_3]} [\Sigma, \Delta]$) does not.
This term must be removed since it's $O(h_3^{-1})$, which is the
leading order in the new terms that are generated by the RG step.
However, this term can be ignored since another unitary
transformation can eliminate it at the expense of only
$O(h_3^{-2})$ terms:
\begin{align}
  H' &\rightarrow H'' = U^\dagger H U \nonumber\\
     &= h_3 \sigma^{[\mu_3]} + \Delta + \frac{1}{2h_3} \sigma^{[\mu_3]} \Sigma^2 + O(h_3^{-2}) \label{eq:H''}
\end{align}
The second approximation (\eqnref{eq:offdiagError}) is used to to
claim that the last term in \eqnref{eq:H'} is indeed
$O(h_3^{-1})$.
%, and can be removed at the expense of only $O(h_3^{-2})$ terms.

Thus, $\sigma^{[\mu_3]}$ is a LIOM of $H''$ at order $O(h_3^{-2})$
if we can assume \eqnref{eq:wolffError} and \ref{eq:offdiagError}.
In \figref{fig:SBRG approx} we show that these appear to be safe
and controlled approximations for sufficiently large $\Gamma$ in
the critical phase. (The approximations are even better in the
spin glass phase.) The first approximation
(\eqnref{eq:wolffError}) gets better under RG flow, while the
second (\eqnref{eq:offdiagError}) does not show a clear trend. If
the second approximation is actually getting worse under RG, this
might suggest that the critical phase in the XYZ model is actually
thermal.
% Specifically, we show that $|\Sigma|/h_3 \sim e^{-\Gamma}$ and $\max [\Sigma, \Delta]/h_3 \sim e^{-\Gamma}$ are both controlled by large disorder $\Gamma$; and both assumptions appear to get better as the RG flows.
% If we wanted to go to a higher order perturbative order, such as $O(h_3^{-3})$, then we would need to make additional approximations, such as
%   $\max_{\Delta_0 \in \Delta, [[\Sigma,\Delta],\Delta_0] \neq 0} |\Delta_0| &\ll h_3$.

Now that $\sigma^{[\mu_3]}$ commutes with the Hamiltonian, we may
want to perform a change of basis $\sigma^{[\mu_3]} \rightarrow
\tau^3_j$. Any integer $j$ can be used as long as it hasn't been
used before. E.g. $j$ could be chosen to indicate the RG step
number, or the rough position of $\sigma^{[\mu_3]}$ on the
lattice. In practice, it is convenient to use the same Hilbert
space for $\sigma$ and $\tau$. This can be done by rotating
$\sigma^{[\mu_3]} \rightarrow \sigma^3_j$ using a unitary
transformation composed of one or two $C_4$ transformations
$\exp\left(\frac{\ii\pi}{4}\sigma^{[\mu]}\right)$ as described in
\cite{You:2015sb}.

\subsection{Edwards-Anderson Correlator via SBRG}
\label{sec:SBRG Anderson}

Given a Hamiltonian $H$, SBRG calculates the Schrieffer-Wolff
transformation $S_\text{tot} = \prod S_j$ (\eqnref{eq:wolff S})
that rotates (\eqnref{eq:wolff H}) the Hamiltonian into a basis
with LIOM $\sigma^{[\mu_3]}_j$. Before applying $S_\text{tot}$,
$\sigma^{[\mu_3]}_j$ are only approximate LIOM of $H$.

We want to calculate the energy averaged Edwards-Anderson
correlator of an operator $Q$:
\begin{equation}
\label{eq:anderson 2}
  \text{avg} \braket{Q}^2 \equiv \frac{1}{N_E} \sum_E \braket{E | Q | E}^2
\end{equation}
To do this, we will apply the Schrieffer-Wolff transformation to
both $\sigma^{[\mu]}$ and $\ket{E}$:
\begin{align}
  H        \rightarrow \tilde{H}
          &\equiv S_\text{tot}^\dagger H S_\text{tot} \label{eq:wolff H} \\
  \ket{E}  \rightarrow \ket{\tilde{E}}
          &\equiv      S_\text{tot}^\dagger \ket{E} \\
  Q        \rightarrow \tilde{Q}
          &\equiv      S_\text{tot}^\dagger Q S_\text{tot} \label{eq:wolff Q}\\
          &=           \sum_{[\mu]} q_{[\mu]} \sigma^{[\mu]} \nonumber
\end{align}
for some set of coefficients $q_{[\mu]}$. Now every
$\sigma^{[\mu_3]}_j$ takes a definite value for every energy
eigenstate $\ket{\tilde{E}}$ of $\tilde{H}$: $\sigma^{[\mu_3]}_j =
\pm 1$. This implies that $\braket{\tilde{E} | \sigma^{[\mu]} |
\tilde{E}} = \pm 1$ if $\sigma^{[\mu]}$ commutes with all
$\sigma^{[\mu_3]}_j$, otherwise it's zero. Thus
\begin{align}
  \text{avg} \braket{Q}^2
    &= \frac{1}{N_E} \sum_E \braket{\tilde{E} | \tilde{Q} | \tilde{E}}^2 \\
    &= \frac{1}{N_E} \sum_E \sum_{[\mu]} \braket{\tilde{E} | q_{[\mu]} \sigma^{[\mu]} | \tilde{E}}^2 \nonumber\\
    &= \sum_{[\mu]}
        \left\{\begin{array}{ll}
           q_{[\mu]}^2 & \sigma^{[\mu]} \text{ commutes with all } \sigma^{[\mu_3]}_j \\
           0           & \text{otherwise}
         \end{array}\right. \nonumber
\end{align}
In the second line, we were able to pull the $\sum_{[\mu]}$ out of
$\braket{\tilde{E} | \cdots | \tilde{E}}^2$ because cross terms
cancel after the energy average. Thus, the energy averaged
Edwards-Anderson correlator can be calculated from the
coefficients $q_{[\mu]}$ of $\tilde{Q}$ (\eqnref{eq:wolff Q}).

\subsection{Algorithm and Implementation Details}
\label{sec:SBRG details}

Unfortunately, the Schrieffer-Wolff transformation generates an
exponentially large number of terms in $\Sigma^2$ in
\eqnref{eq:H''} and $\tilde{Q}$ in \eqnref{eq:wolff Q}. Thus, one
must drop terms. In this work, for each Schrieffer-Wolff
transformation, we keep the 256 largest additional terms in
$\Sigma^2$ and 32 largest additional terms in $\tilde{Q}$.

Most operators $\sigma^{[\mu]}$ in the Hamiltonian will be local.
Thus, it is important to make use of this locality when
implementing SBRG. With this optimization, the CPU time for an
SBRG simulation of a Hamiltonian in an MBL phase scales linearly (up to log corrections)
with system size.

For this work, each $L=256$ disorder realization data point took a
couple minutes of CPU time. However, reasonable data is achievable
with only 1/8th as many additional terms in $\Sigma^2$ and
$\tilde{Q}$, for which a simulation only takes several seconds.
The SBRG data used in this paper was calculated in roughly two weeks on
a quad-core i7-3720QM underclocked to 2GHz. The SBRG data included
1024 disorder realizations for each data point. However,
reasonable data was attainable with less ($\sim100$) disorder
realizations and additional terms, for which only a few hours of
simulation time is necessary.

For this work, the Haskell programming language was used to
implement SBRG in roughly 1000 lines of code, and the implementation has been published to github \cite{Slagle:github}. Haskell was chosen
because it could produce fast code (potentially faster than C++
due to very efficient garbage collection) while requiring the
smallest amount of developer time compared to other languages.
Development time was minimized using Haskell because Haskell is
very good at dealing with complicated data structures, which were
necessary for making use of the locality of operators mentioned
above. This is a result of Haskell's strong type system, automatic
garbage collection, functional programming paradigm, and high
amount of expressiveness.

Random numbers were generated using a combined linear congruential
generator (System.Random.StdGen in Haskell) with period $2^{61}$,
which is much larger than the number of random numbers used in
this work, which was roughly $2^{29}$.

\section{XYZ Model Details}

\subsection{Disorder Distribution}
\label{sec:disorder distribution}

SBRG is a numerical method which relies on large disorder for
accuracy. For this reason, it is important that we choose a
disorder distribution with very large randomness. For the XY
($J_z=0$) and XXZ ($J_{i,x}=J_{i,y}$) spin chains, it has been
shown that with large disorder, $J_{i,\mu}$ flows to a power law
distribution \cite{Fisher:1992is,Fisher:1994he} with a probability
distribution function (PDF) equal to
\begin{equation}
  \mathrm{PDF}(J_{i,\mu}) = \frac{1}{\Gamma J_{i,\mu}} \left(\frac{J_{i,\mu}}{J_{\mu}}\right)^{1/\Gamma} \quad,\quad J_{i,\mu} > 0 \label{eq:Jdist}
\end{equation}
where $\Gamma$ controls the strength of the disorder, with larger
$\Gamma$ corresponding to stronger disorder. It's useful
to define $\beta_{i,\mu} \equiv -\ln \frac{J_{i,\mu}}{J_\mu} \ge
0$. With this definition, $\beta_{i,\mu}$ follows an exponential
distribution with a mean and standard deviation (as in \eqnref{eq:stdJ}) equal to $\Gamma$:
\begin{equation}
  \mathrm{PDF}(\beta_{i,\mu}) = \Gamma^{-1} e^{-\beta_{i,\mu}/\Gamma} \quad,\quad \beta_{i,\mu} > 0
\end{equation}
It's also useful to define $\J_{i,\mu}$ by $J_{i,\mu} \equiv
\J_{i,\mu}^\Gamma$. With this definition, $\J_{i,\mu}$ is
uniformly distributed in $[0,\J_\mu]$ where $J_\mu \equiv
\J_\mu^\Gamma$:
\begin{equation}
\label{eq:PDF tilde J}
  \mathrm{PDF}(\J_{i,\mu}) = \J_{\mu}^{-1} \quad,\quad 0 < \J_{i,\mu} < \J_\mu
\end{equation}

We therefore let $J_{i,\mu}$ follow the above (equivalent)
distributions so that we can effectively probe closer to the
infinite system size limit while using the same physical system
size. (We assume that the disorder becomes stronger under RG in
the large disorder limit.) We will show results for $\Gamma = 4$
because this is the smallest $\Gamma$ for which SBRG agrees well
with exact diagonalization on small latices (data not shown in
this paper), and for which our approximations appear to be safe
(see \appref{sec:SBRG Approximations} and \ref{sec:ED}, and \secref{sec: MBL vs
thermal}).

\subsection{Tuning Parameters}
\label{sec:tuning parameters}

We use $\J_\mu$ as tuning parameters instead of $J_\mu$ because
$\J_\mu$ are better behaved at large disorder $\Gamma$. For
example, our phase diagram in \figref{fig:phase diagram} would
depend strongly on $\Gamma$ if $J_\mu$ were used as tuning
parameters instead of $\J_\mu$. However, when $\J_\mu$ are used as
tuning parameters, the phase diagram has little dependence on
$\Gamma$ when $\Gamma$ is large; and this allows the large
disorder limit to be well defined.

A simple calculation makes this more transparent. If $P(J_{i,\mu}
> J_{i,\nu})$ is the probability that $J_{i,\mu} > J_{i,\nu}$,
then $P(J_{i,\mu} > J_{i,\nu}) \rightarrow \frac{1}{2}$ as $\Gamma
\rightarrow \infty$ if $J_\mu$ is held constant. Thus large
disorder effectively washes out the differences between different
$J_\mu$ and the $J_1 = J_2 = J_3$ point effectively dominates the
phase diagram if $J_\mu$ is held constant as $\Gamma \rightarrow
\infty$. However, if we define $J_\mu \equiv \J_\mu^\Gamma$ and
hold $\J_\mu$ constant instead, then $P(J_{i,\mu} > J_{i,\nu})$ is
independent of $\Gamma$. This is because
\begin{align*}
  P(J_{i,\mu} > J_{i,\nu})
    &= P(\J_{i,\mu} > \J_{i,\nu}) \text{ since } J_{i,\mu} \equiv \J_{i,\mu}^\Gamma \\
    &= \begin{cases} \;\;\;\frac{r}{2}  & \text{if } r<1 \\
                       1 - \frac{1}{2r} & \text{if } r>1 \end{cases} \text{ by \eqnref{eq:PDF tilde J}} \\
    &  \quad\quad\text{where } r = \frac{\J_\mu}{\J_\nu} = \left( \frac{J_\mu}{J_\nu} \right)^{1/\Gamma}
\end{align*}

\section{Exact Diagonalization Study}
\label{sec:ED}

\begin{figure}[thbp]
\includegraphics[width=\columnwidth]{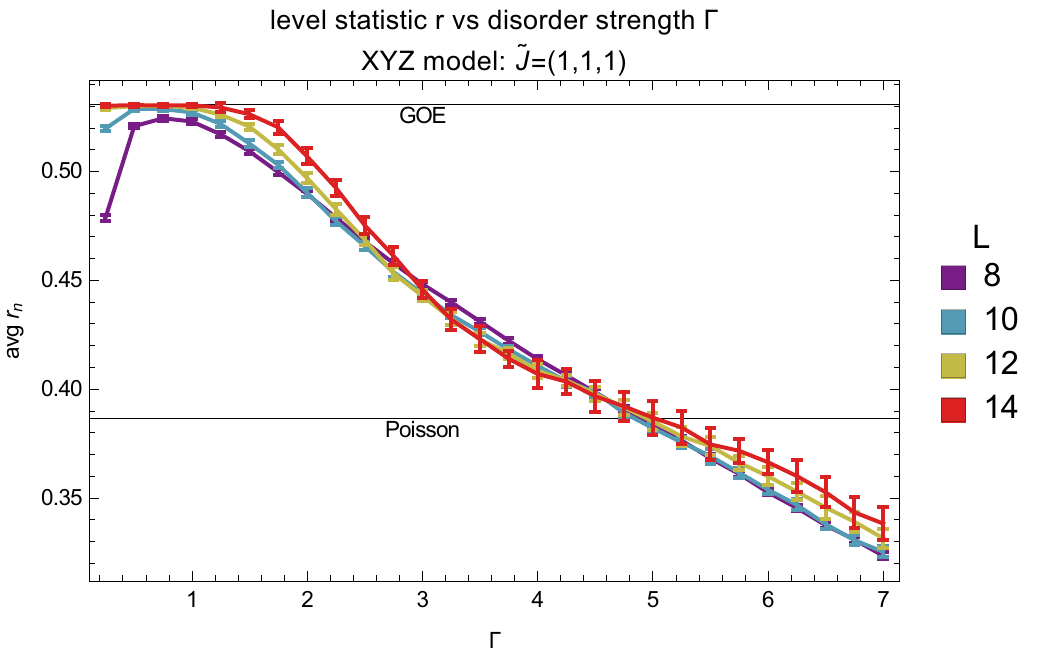}
\caption{Disorder and energy averaged level statistic $r$ (\eqnref{eq:levStat}) vs disorder strength $\Gamma$ for different system sizes $L$ in the XYZ spin chain with $\J_x = \J_y = \J_z = 1$.
With weak disorder, the system approaches GOE level statistics with increasing system size, which indicates that the system is thermal.
As the disorder strength increases, the level statistic drops below the Poisson level statistic, which suggests that the system is not thermal with strong disorder.
% We also show a $\J = (0,0,1)$ line as a simple example of another system with level statistics below the Poisson level statistic.
\footError}
\label{fig:levStat}
\end{figure}

We now study the critical point $\J_x = \J_y = \J_z = 1$ using exact diagonalization to check that the critical phase is not thermal.
In \figref{fig:levStat} we show the level statistic $r$ of the XYZ model vs disorder strength $\Gamma$.
For each disorder realization, the level statistic $r_n^{(a)}$ is defined as a ratio of level spacings $\delta_n^{(a)}$ \cite{Huse:2007is} in a given symmetry sector~$(a)$:
\begin{align}
  r_n^{(a)}       &\equiv \frac{\min(\delta_n^{(a)},\delta_{n+1}^{(a)})}{\max(\delta_n^{(a)},\delta_{n+1}^{(a)})} \label{eq:levStat}\\
  \delta_n^{(a)}  &\equiv E_n^{(a)} - E_{n-1}^{(a)} \nonumber\\
  E_{n+1}^{(a)}   &\geq E_n^{(a)} \nonumber
\end{align}
For our model, there are four symmetry sectors $(a)$ which are labeled by the eigenvalues of the symmetry operators of our model: $\prod_i \sigma_i^x$ and $\prod_i \sigma_i^y$.
In \figref{fig:levStat} we average over disorder realizations, level spacings $n$, and symmetry sectors $(a)$.
We find that with weak disorder $\Gamma$, the level statistic approaches (with increasing system size) the GOE (Gaussian orthogonal ensemble) level statistic $r_\text{GOE} \approx 0.53$, which is expected for a thermal system.
As the disorder strength increases, the level statistic drops below the Poisson level statistic $r_\text{Poisson} \approx 0.39$, which suggests that the system is not thermal.

The level statistic continues below the Poisson level statistic because we use a power law distribution of coefficients in the Hamiltonian (\eqnref{eq:Jdist}), instead of a uniform or Gaussian distribution.
This can be understood most easily in the $\J_x = \J_y = 0$ and $\J_z = 1$ limit (see \figref{fig:levStat}), where the Hamiltonian is already diagonal.
When $\Gamma = 1$, we exactly reproduce the Poisson level statistic.
However, a simple numerical calculation shows that as $\Gamma$ increases, the level statistic decays to zero as a power law with increasing $\Gamma$: $r \sim 1/\Gamma$.

\begin{figure}[thbp]
\includegraphics[width=\columnwidth]{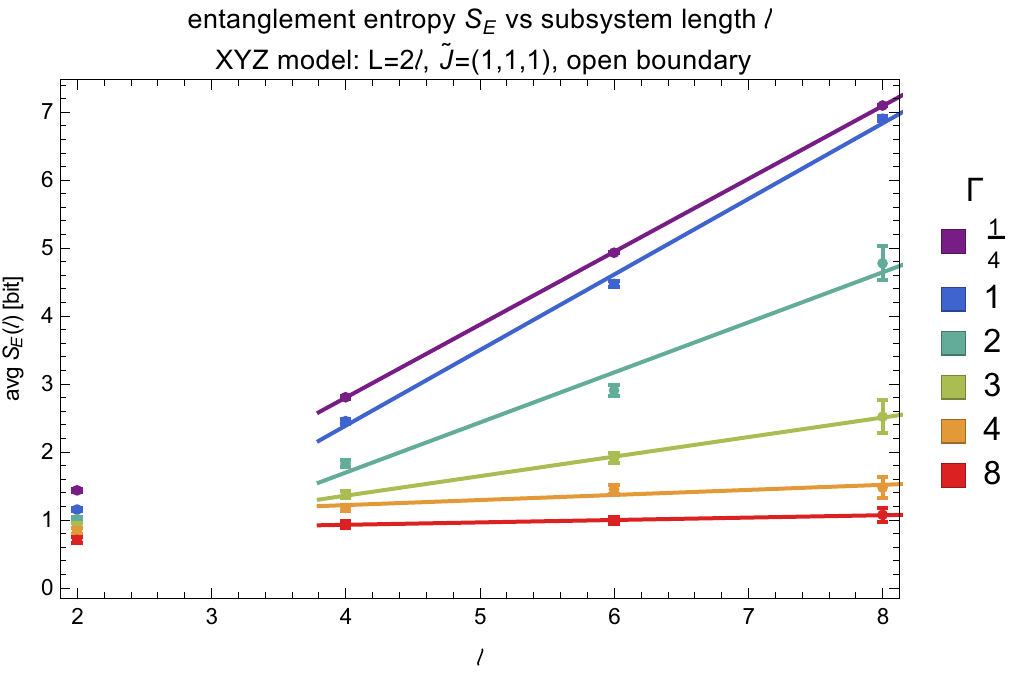}
\caption{Disorder and energy averaged entanglement entropy $\SE$ [bit] ($\equiv \SE / \ln2$) (\eqnref{eq:S_E}) vs subsystem length $\ell$ for different disorder strengths $\Gamma$ in the XYZ spin chain
with $\J_x = \J_y = \J_z 1$ and system size $L=2\ell$ with open boundary conditions.
Each entanglement subsystem splits the system in half.
We only show even $\ell$ in order to avoid an even-odd effect.
As the disorder strength $\Gamma$ increases, the slope of the data decreases to roughly zero.
This suggests a transition from a thermal phase with volume law entanglement to a non-thermal phase. \footError}
\label{fig:ED entanglement}
\end{figure}

In \figref{fig:ED entanglement} we show the how the entanglement scales with subsystem size.
When the disorder is weak, the system displays a volume law entanglement, as expected for a thermal system.
But for strong disorder, the entanglement increases very slowly with subsystem size, which suggests that the strong disorder results in either a MBL phase with area law entanglement or a marginal MBL phase with log-law entanglement.
(In \figref{fig:entanglement} we SBRG and large system sizes to show that the entanglement follows a log-law.)
Again we see that for weak disorder $\Gamma$, the critical point is thermal, but with strong disorder the critical point does not appear to be thermal.

It is worth emphasizing that in both ED plots, a disorder strength of $\Gamma = 1$ (which corresponds to uniformly distributed random coefficients $J_i$) was not sufficient to prevent thermalization of the XYZ model's critical point.
Therefore, considering disordered systems with only Gaussian (with $\Gamma \approx 1.1$ via \eqnref{eq:stdJ}) or uniform disorder distributions may not always be sufficient if one is interested in observing marginal MBL critical phases.
That is, similar to the XYZ model's critical point, other models may also require large $\Gamma$ (defined by \eqnref{eq:stdJ}) in order to prevent thermalizaion.

\section{Spin Glass Entanglement Entropy Calculation}
\label{sec:SG EE}

\begin{figure}[thbp]
\includegraphics[width=\textwidth]{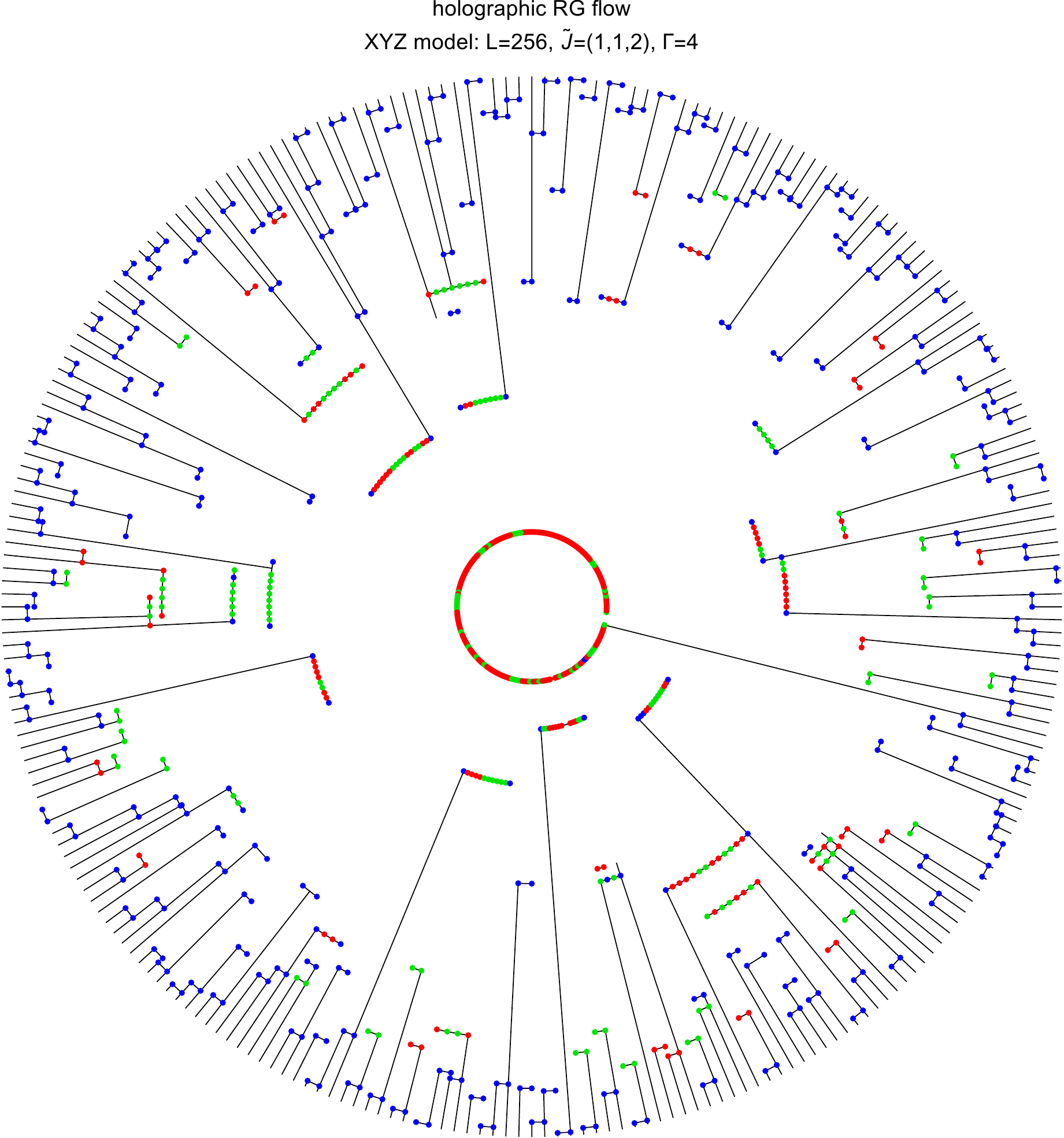}
\caption{ Approximate RG flow and LIOM for the XYZ spin chain in
the spin glass phase with $\J = (1,1,2)$. (See the caption of
\figref{fig:holography critical}.) Because we are in the spin
glass phase, the LIOM are dominated by blue $\sigma^z \sigma^z$
operators. The very large integral of motion in the center is just
the $Z_2$ spin flip symmetry $\prod_i \sigma^x_i$, which is a
little messy in the figure because it has been multiplied by some
of the other LIOM to obtain an operator which is still an integral
of motion. This center operator is essential for the spin glass
properties of this phase: finite long range Edwards-Anderson
correlators (\figref{fig:anderson}) and finite long range mutual
information (\figref{fig:MI}). }\label{fig:holography SG}
\end{figure}

Here we calculate the entanglement entropy of a subsystem $A$ in
the spin glass phase using the stabilizer rank algorithm in
\cite{You:2015sb}. Deep in the spin glass phase where $J^z$
dominates, the LIOM are given by: \begin{align} \label{eq:SG LIOM}
  \tau_i^z &= \sigma_i^z \sigma_{i+1}^z \text{ with } i=1,2,\ldots L-1 \\
  \tau_L^z &= \prod_{i=1}^L \begin{cases} \sigma^x & \text{if } J_x > J_y \\
                          \sigma^y & \text{if } J_y > J_x \end{cases} \nonumber
\end{align}
Assuming $i<j$. We can see that in this case indeed $\sigma_i^z
\sigma_j^z = \prod_{k=i}^{j-1} \tau_k^z$ is a product of the LIOM.
Only three of the LIOM will be cut by the subsystem $A$, which we
will take to be the sites $i,i+1,\ldots j$ where $j = i-1+\ell$.
These LIOM are $\tau_{i-1}^z, \tau_j^z, \tau_L^z$. We then ``trace
out degrees of freedom not in $A$" by removing $\sigma$-matrices
not in $A$:
\begin{align*}
      \tau_{i-1}^z &= \sigma^z_{i-1} \sigma^z_i &     \tau_j^z &= \sigma^z_j \sigma^z_{j+1} &     \tau_L^z &= \prod_{i'} \sigma^x_{i'} \\
                   &\downarrow                  &              &\downarrow                  &              &\downarrow \\
  \hat\tau_{i-1}^z &\rightarrow      \sigma^z_i & \hat\tau_j^z &\rightarrow  \sigma^z_j     & \hat\tau_L^z &\rightarrow \prod_{i'=i}^j \sigma^x_{i'}
\end{align*}
We then consider the 3x3 anticommutivity matrix of these three
$\hat\tau$ operators:
\[ M = \mat{0 & 0 & 1 \\ 0 & 0 & 1 \\ 1 & 1 & 0} \]
where a $1$ denotes anticommutivity while a $0$ denotes commutivity.
The entanglement entropy $\SE$ is then given by
\begin{align}
\label{eq:H_eff EE}
\SE &= \frac{\ln2}{2} \text{rank} M \text{ (over $\dsZ_2$)} \\
           &= \ln2 \nonumber
\end{align}
where the matrix rank is calculated over the field $\dsZ_2$. When
one isn't deep in the spin glass phase, the LIOM become more
complicated (as in \figref{fig:holography SG}); more LIOM are cut by $A$; and the entanglement
entropy increases slightly due to additional boundary
contributions which don't depend significantly on the size of the
subsystem $A$.

%\section{Algorithm Challenge: Missing Stabilizers}

\newpage
\onecolumngrid

\begin{figure}[thbp]
\includegraphics[width=\textwidth]{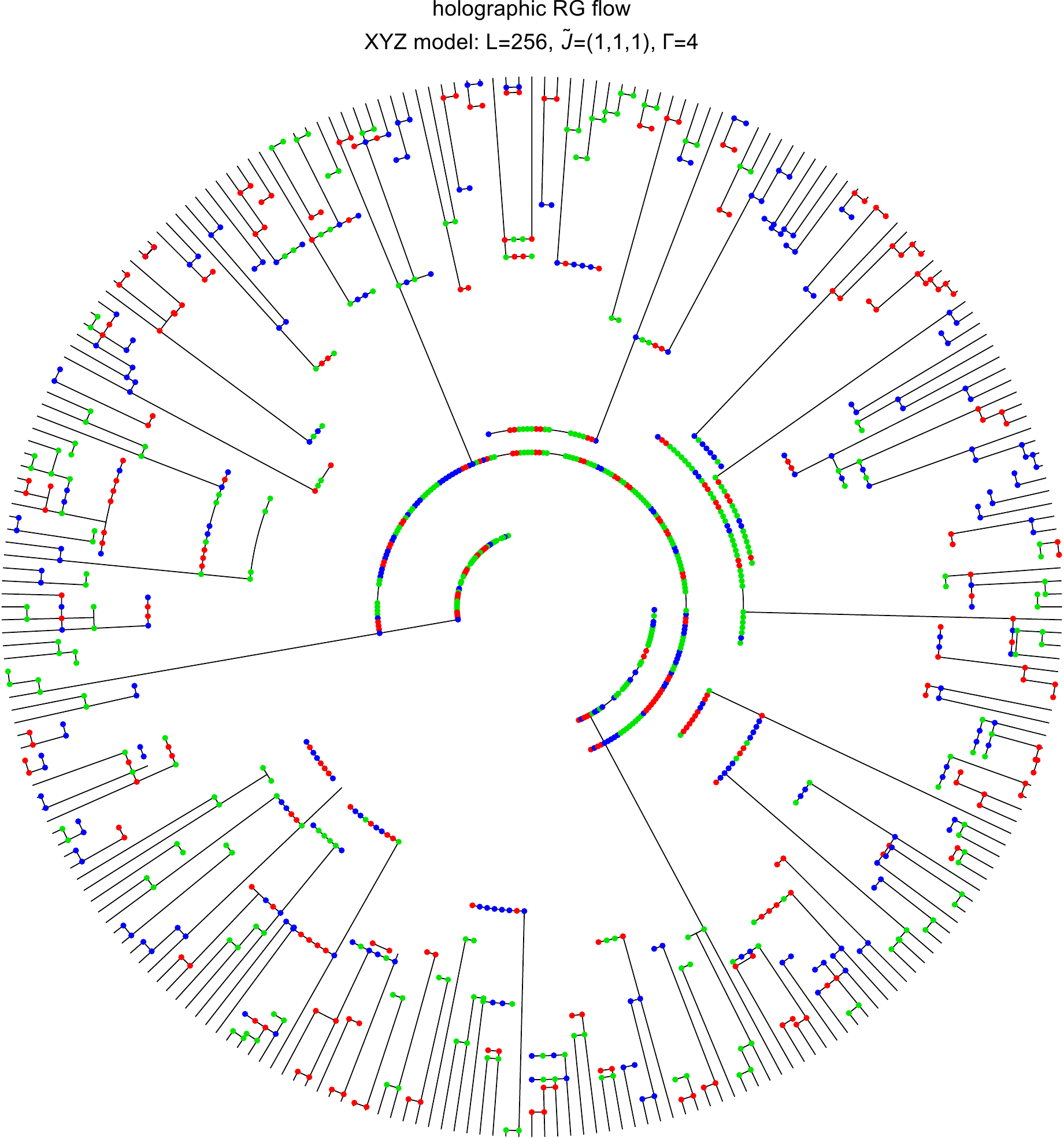}
\caption{ Approximate RG flow and LIOM for the XYZ spin chain at
the critical point $\J_x = \J_y = \J_z = 1$. The initial
Hamiltonian starts at the boundary of the disk, with a spin living
at the end of each radial line at the edge of the disk. SBRG then
performs an RG which identifies a LIOM at each step of the RG (see
\appref{sec:SBRG Approximations}). In the limit of large disorder,
the LIOM can be approximated as a product of $\sigma^x$ (red),
$\sigma^y$ (green), and $\sigma^z$ (blue) matrices (dots connected
by an arc line in the figure). Each LIOM in the figure also marks
the end of a radial line to denote that a degree of freedom has
been diagonalized (cf. ''integrated out``) at this RG step.
Because we are in a marginal MBL phase, LIOM exist with sizes at
every length scale, which results in the critical properties of
this phase: diverging entanglement entropy
(\figref{fig:entanglement}), power-law Edwards-Anderson
correlators (\figref{fig:anderson}), and long power-law range
mutual information (\figref{fig:MI}). }\label{fig:holography
critical}
\end{figure}

\end{document}